\documentclass[useAMS,usenatbib]{mn2e}

\usepackage{wasysym}

\usepackage{graphicx}
\usepackage{subfigure}

\title[Using \textup{MOST} to reveal the secrets of CV Ser]{Using \textit{MOST} to reveal the secrets of the mischievous Wolf-Rayet binary CV Ser}
\author[A. David-Uraz et al.]{Alexandre David-Uraz,$^{1}$\thanks{E-mail:
alexandre@astro.umontreal.ca (ADU); moffat@astro.umontreal.ca (AFJM)}  Anthony F. J.
Moffat,$^{1}$\footnotemark[1] Andr\'{e}-Nicolas Chen\'{e},$^{2}$ $^{3}$ 
\newauthor 
Jason F. Rowe,$^{4}$ Nicholas Lange,$^{5}$ David B. Guenther,$^{6}$ Rainer Kuschnig,$^{7}$ $^{8}$
\newauthor
Jaymie M. Matthews,$^{8}$ Slavek M. Rucinski,$^{9}$ Dimitar Sasselov$^{10}$ and
\newauthor
Werner W. Weiss$^{7}$\\
$^{1}$D\'{e}partement de physique and Centre de Recherche en Astrophysique du Qu\'{e}bec, Universit\'{e} de Montr\'{e}al, C.P. 6128, Succursale\\
Centre-Ville, Montr\'{e}al, QC H3C 3J7, Canada\\
$^{2}$Departamento de Astronom\'i{}a, Universidad de Concepci\'o{}n, Casilla 160-C, Concepci\'o{}n, Chile\\
$^{3}$Departamento de F\'i{}sica y Astronom\'i{}a, Facultad de Ciencias, Universidad de Valpara\'i{}so, Av. Gran Breta\~n{}a 1111, Playa Ancha, \\
Casilla 5030, Valpara\'i{}so, Chile\\
$^{4}$NASA Ames Research Center, Moffett Field, CA 94035, USA\\
$^{5}$Department of Physics and Astronomy, University of Victoria, Elliott Building, 3800 Finnerty Road, Victoria, BC V8P 1A1 Canada\\
$^{6}$Institute for Computational Astrophysics, Department of Astronomy and Physics, Saint Mary's University, Halifax,\\
NS B3H 3C3, Canada\\
$^{7}$Universit\"{a}t Wien, Institut f\"{u}r Astronomie, T\"{u}rkenschanzstrasse 17, A-1180 Wien, Austria\\
$^{8}$Department of Physics \& Astronomy, University of British Columbia, 6224 Agricultural Road, Vancouver, BC V6T 1Z1, Canada\\
$^{9}$Department of Astronomy and Astrophysics, University of Toronto, 50 St George Street, Toronto, Ontario M5S 3H4, Canada\\
$^{10}$Harvard-Smithsonian Center for Astrophysics, 60 Garden Street, Cambridge, MA 02138, USA}
\begin{document}

\date{Submitted 2011 August 25}

\pagerange{\pageref{firstpage}--\pageref{lastpage}} \pubyear{2011}

\maketitle

\label{firstpage}

\begin{abstract}
The WR binary CV Serpentis (= WR113, WC8d + O8-9IV) has been a source of mystery since it was shown that 
its atmospheric eclipses change with time over decades, in addition to its 
sporadic dust production. The first high-precision time-dependent 
photometric observations obtained with the MOST space telescope in 2009 show 
two \textit{consecutive} eclipses over the 29d orbit, with varying depths. A subsequent 
MOST run in 2010 showed a seemingly asymmetric eclipse profile. In order to help 
make sense of these observations, parallel optical spectroscopy was obtained 
from the Mont Megantic Observatory (2009, 2010) and from the Dominion Astrophysical 
Observatory (2009).  Assuming these depth variations are entirely due to electron scattering in a $\beta$-law wind,
an unprecedented 62\% increase in $\dot{M}$ is observed over one
orbital period.  Alternatively, no change in mass-loss rate would be required if a relatively small fraction of the carbon ions in the
wind globally recombined and coaggulated to form carbon dust grains. However, it remains a mystery as to how this could occur.
There also seems to be evidence for the presence of corotating interaction regions (CIR) in the WR wind: a 
CIR-like signature is found in the light curves, implying a potential rotation period for the WR star of 1.6 d.  Finally, a new
circular orbit is derived, along with constraints for the wind collision.
\end{abstract}

\begin{keywords}
binaries: eclipsing -- stars: mass-loss -- stars: winds, outflows -- stars: Wolf-Rayet.
\end{keywords}

\section{Introduction}

Ever since \citet{1972A&A....20..333A} showed an IR excess in some WC9 stars, it has been known
that certain late-type WCs produce dust.  This dust is composed of amorphous carbon
grains \citep{1987A&A...182...91W} and its formation is also favored in suitable WC + O binaries
(i.e. binaries with large enough separations with respect to each star's luminosity, so that the ionizing flux from the stars cannot prevent
dust formation), presumably
because of the high densities attained in the shocked region between the colliding winds.

CV Ser (= HD 168206 = WR113, $\alpha$ (J2000.0) = 18:19:07.36, $\delta$ (J2000.0) = -11:37:59.2, v = 9.2) 
is a long-studied WC8d+O8-9IV spectroscopic binary \citep{1945ApJ...101..356H} with atmospheric eclipses and a 29.704d period
\citep{1996RMxAC...5..100N}.  Following
the first published light-curve by \citet{1949PZ......7...36G}, various other light curves have shown different eclipse
depths or even no eclipse whatsoever (e.g. \citealt{1963ApJ...137.1080H, 1970AcA....20...13S, 1970ApJ...160L.185K, 1977Obs...97....76W, 1985Ap&SS.109...57L}).
Different explanations were given, including the possibility that the authors had used the wrong orbital period \citep{1971A&A....11..407C}.
However, even when no eclipses were found in the optical continuum, \citet{1972SvA....15..955C} showed that the system was still eclipsing
in the $\lambda 4653$ emission line (confirmed by \citealt{1972PASP...84..635M}). Since then, two MOST 
(Microvariability and Oscillations of STars; see below) runs conducted in 2009 and 2010
also show varying depths, possibly implying a varying mass-loss rate.

CV Ser was also shown to produce dust \citep{1975A&A....40..291C}.  This phenomenon was seen as a plausible explanation for the variation in its 
optical eclipses. It has since been classified as a persistent dust producer \citep{1995IAUS..163..335W}.

Several studies have been carried out to refine the orbital solution of CV Ser.  Most of them found a quasi circular orbit (e.g. \citealt{1981ApJ...245..195M})
but more recently, \citet{1996RMxAC...5..100N} have cast some doubt on that result, finding an eccentricity of 0.19.

Initially, the goal of this project was to monitor stochastic short-term absorption features in CV Ser's light curve 
involving light from the orbiting O star in order to try and link 
them to clumps in the WR wind.
Because of its high sensitivity, the MOST space telescope seemed like the perfect instrument to carry out
this research. The system was observed in 2009 for more than a complete orbital cycle, since the anticipated absorption associated with the clumps should vary
with the orbital phase, depending on the part of the WR wind illuminated by the O star along the line of sight (see Fig.~1).

\begin{figure*}
\includegraphics[width=5.0in]{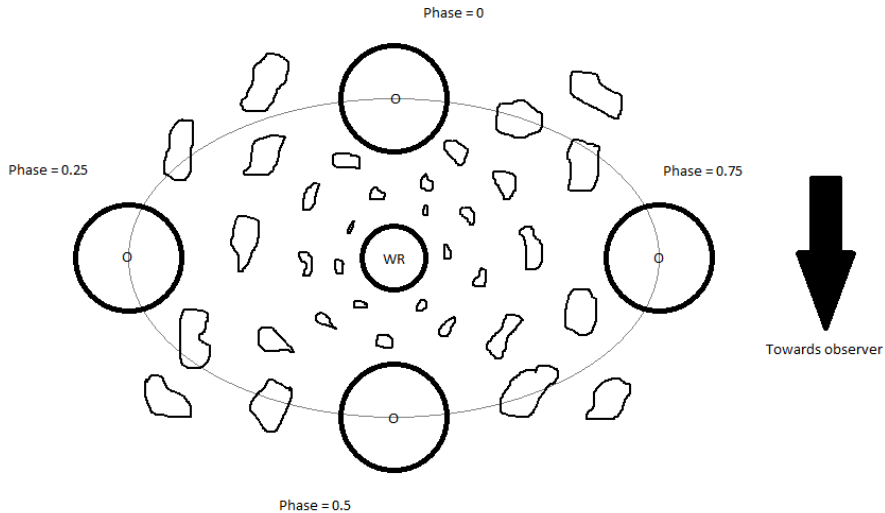}
\caption{Intuitive cartoon to show how it is expected that the absorption by the clumps along the line of sight should depend on the orbital phase.}
\label{fig:1}
\end{figure*}

If the clumps produce observable absorption throughout the orbit, one would expect to find shorter and deeper variations near $\phi = 0$ (WR 
at inferior conjunction) and almost no random
variations (unless one of the stars undergoes intrinsic variability) around $\phi = 0.5$.  Therefore, the scatter of the residuals from a given
light curve fit 
should vary with phase, being larger around $\phi = 0$.  These variations would be a few mmag deep, as expected from the typical relative
density enhancement in clumps.

Section~\ref{sec:obs} will briefly summarize the observations (both photometric and spectroscopic) of CV Ser taken for this study. The specific results
for each observing run are presented in section~\ref{sec:anal}, and then are briefly discussed in section~\ref{sec:disc}.

\section{Observations}\label{sec:obs}

\subsection{Optical photometry}

Two photometry runs (2009 and 2010) were carried out using the MOST space telescope \citep{2003PASP..115.1023W}, a 150-mm aperture 
Rumak-Maksutov telescope with, initially, 2 CCDs 
(1 for science and 1 for tracking; however, the tracking CCD was lost in 2007 after being hit by a cosmic ray and tracking and science images are now both
obtained on the same CCD) on a polar, Sun-synchronous, low-Earth 101-minute orbit.  MOST is used for optical photometry, with a wideband filter (3000 
\AA{}, centered around 5250 \AA{}).

The first run was obtained contiguously from 2009 June 27 to 2009 August 11.  This 45-day observation encompassed 2 eclipses of CV Ser, with
millimagnitude precision.  Most data points are 0.5 min apart, except when the satellite went over the South Atlantic Anomaly (SAA).  Part of the
run was also shared with another target, giving an overall somewhat inhomogeneously distributed data set.  One particularity of this light curve is the
difference of depth between both eclipses, which will be discussed in more detail in the next section.  

Because of this interesting behavior, CV Ser was observed once again with MOST the following year, from 2010 June 14 to 2010 July 11, 
in similar conditions.  However, this run covered only one eclipse of CV Ser.

\subsection{Optical spectroscopy}

Spectroscopy was obtained at both Observatoire du Mont-M\'{e}gantic (2009 and 2010) and the Dominion Astrophysical Observatory (2009 only).  Details of these
observations are summarized in Table~\ref{tab:spectro}.

\begin{table*}
\caption{Summary of the spectroscopic observations.}
\label{tab:spectro}
\begin{tabular}{|c|c|l|c|c|c|c|}
\hline
Telescope & Year & Dates & Number of spectra & Spectral range ($\textrm{\AA}$) & Resolution ($\textrm{\AA}$/pixel) & Average SNR \\
\hline
OMM (1.6m) & 2009 & July 5 - August 8 & 145 & 4400-6000 & $\sim 0.64$ & $\sim 150$ \\
DAO (1.8m) & 2009 & July 7 - July 31  & 156 & 5300-6100 & $\sim 0.77$ & $\apprge 100$ \\
OMM (1.6m) & 2010 & June 21 - July 12 & 50  & 4400-6000 & $\sim 0.64$ & $\sim 150$ \\
\hline
\end{tabular}
\end{table*}

Unfortunately, some of the 2010 OMM spectra are full of narrow instrumental defects.  As this effect could not be corrected, these spectra had to be discarded.
The spectra were reduced using IRAF \footnote{IRAF is distributed by the National Optical Astronomy Observatories (NOAO), which is operated by the
Association of Universities for Research in Astronomy, Inc., (AURA) under cooperative agreement with the National Science Foundation (NSF).}.  
All spectra were normalized to the continuum.  An example spectrum from OMM is shown in Fig.~2, revealing the main WC8 spectral features.

\begin{figure}
\includegraphics[width=3.4in]{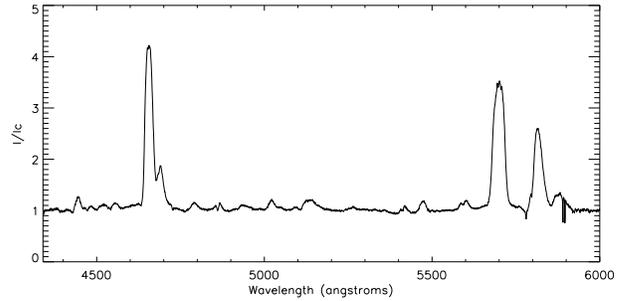}
\caption{Sample spectrum from the 2009 OMM dataset.}
\label{fig:2}
\end{figure}

\section{Analysis and Results}\label{sec:anal}

\subsection{MOST 2009 Dataset}\label{ss:2009phot}

First, without loss of information, the data were binned for each MOST orbit (101 minutes), 
applying 3-sigma clipping for each binned point.  The resulting light curve is shown in Fig.~3a.
A steady linear drift can be seen throughout the light curve (almost 0.01 mag over 45 days).  It did not seem to be due only to the stars and it is well
known that such instrumental drifts occur over long periods of continuous observation with the MOST satellite.
Similar drifts are visible in the light curves of other field stars, but since the slope of these drifts does not vary smoothly from one region to the 
next on the CCD, it is impossible to map them out in an effort to determine what part of the drift of the CV Ser light curve is really instrumental.  The only option
was to fit it out.  Nevertheless, both eclipses had undeniably different depths, unless an unusually deep stochastic dip (e.g. due to a large clump), 
like those seen elsewhere in the light 
curve, happened to occur exactly at phase 0.00.  Not only does such a coincidence seem extremely unlikely, but a stochastic dip would not likely
produce the same eclipse profile due to the global WR wind.  Therefore, we dismiss this possibility as being implausible.

The data were fitted using the model developed in \citet{1996AJ....112.2227L}: the eclipse is atmospheric and occurs as the O star goes behind the WR wind,
its light being Thomson scattered by free electrons in the WR wind.  This assumption is reasonable, because at the considered wavelengths, Thomson
scattering is clearly the dominating process. Compared, for example, to free-free absorption, Thomson scattering's influence on the light curve is
more than one order of magnitude greater in the visible spectrum, even in the densest wind regions (at $R = 2 R_{*}$). Only in the NIR do these processes become
comparable. This is easily verifiable using the appropriate formulae of \citet{1977rpa..book.....T}. A circular orbit is also assumed in this model.

The Lamontagne et al. model is purely geometrical and it yields various parameters of the system, including the orbital
inclination and the mass-loss rate of the WR component.  One of the advantages of this model is that the obtained mass-loss rate is not affected by clumping,
and since it makes simple assumptions, it is fairly easy to use.

For wind velocity laws with integer values of $\beta$, the model produces an analytical formula describing the behavior of the light curve.  In this case we used
a $\beta = 0$ law (i.e. constant expansion velocity), although it is not the closest to reality (at least not in the case of hot stars, where we have
$\beta \sim 0.8$, \citealt{1986A&A...164...86P}). However, as shown in \citet{1996AJ....112.2227L}, both
$\beta = 0$ and $\beta = 1$ velocity laws in practice produce very similar synthetic light curves, and the first case requires fewer free parameters.  
The corresponding formula is

\begin{equation}
   \Delta m = {\Delta m}_{0} + A \left(\frac{\pi/2 + \arcsin \epsilon}{\sqrt{1-{\epsilon}^2}}\right)
\end{equation}

\noindent with $\Delta m_{0}$ a constant, $\epsilon = (\sin i) \cos 2 \pi \phi$, 

\[
   A = \frac{(2.5 \log e) k}{(1 + I_{WR} / I_{O})}\ \textrm{and}\ k = \frac{\alpha \sigma_{e} \dot{M}}{4 \pi m_{p} v_{\infty} a},
\]

\noindent in which $i$ is the orbital inclination, $e$ is Euler's number ($\simeq 2.718$), $I_{WR} / I_{O}$ is the intensity ratio of 
the 2 stars in the observed bandpass, $\alpha \simeq$ 0.5 free electron per baryon mass (since the dominating element is fully ionized helium), 
$\sigma_{e}$ is the Thomson electron-scattering cross section, $\dot{M}$ is the WR mass-loss rate, $m_{p}$ is the proton mass,
$v_{\infty}$ is the terminal wind speed and $a$ is the orbital separation.

\begin{figure*}
\begin{center}
\subfigure[Modified Lamontagne fit to the 2009 MOST light curve.]{\includegraphics[width=3.4in]{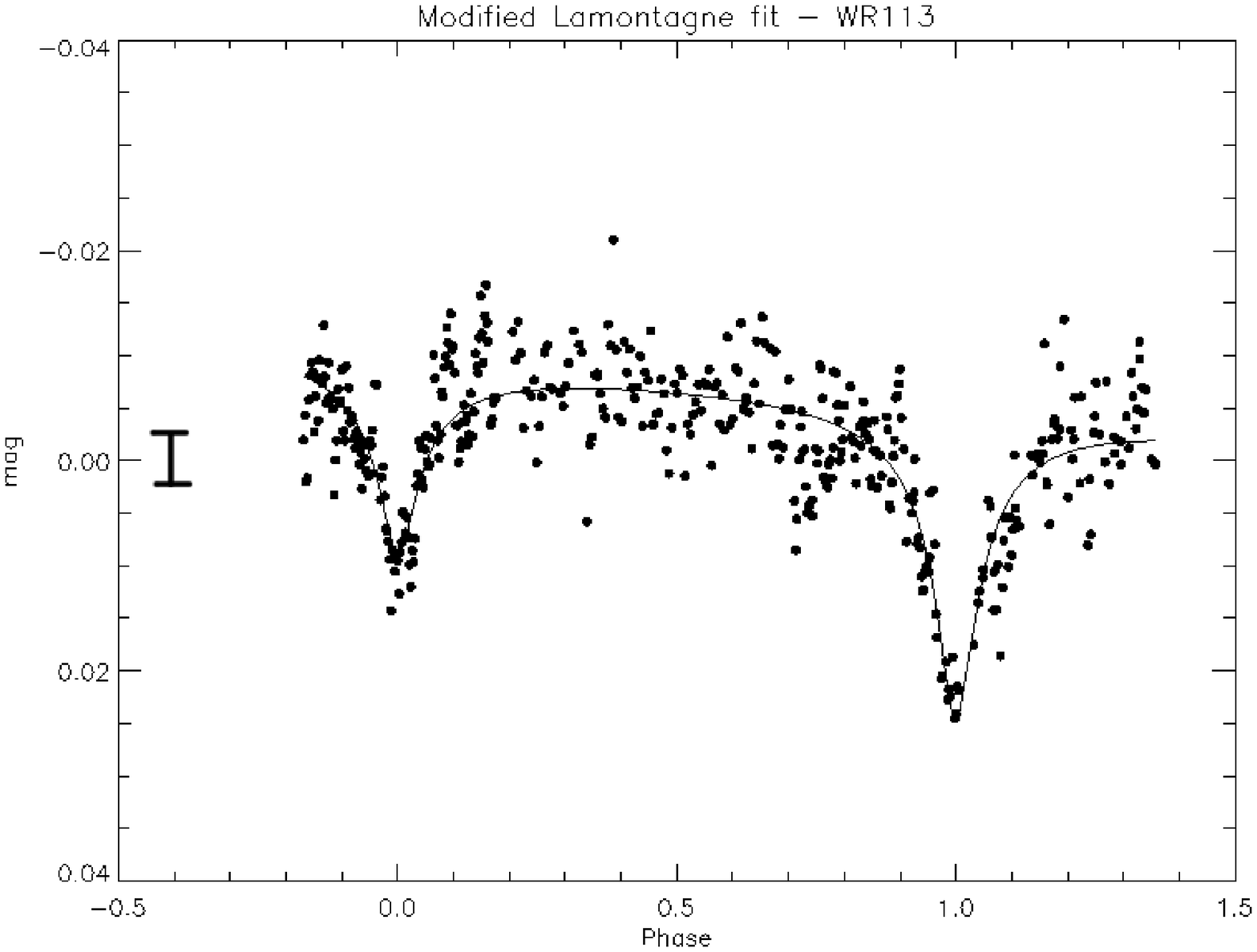}}
\subfigure[Lamontagne fit to the 2010 data.]{\includegraphics[width=3.4in]{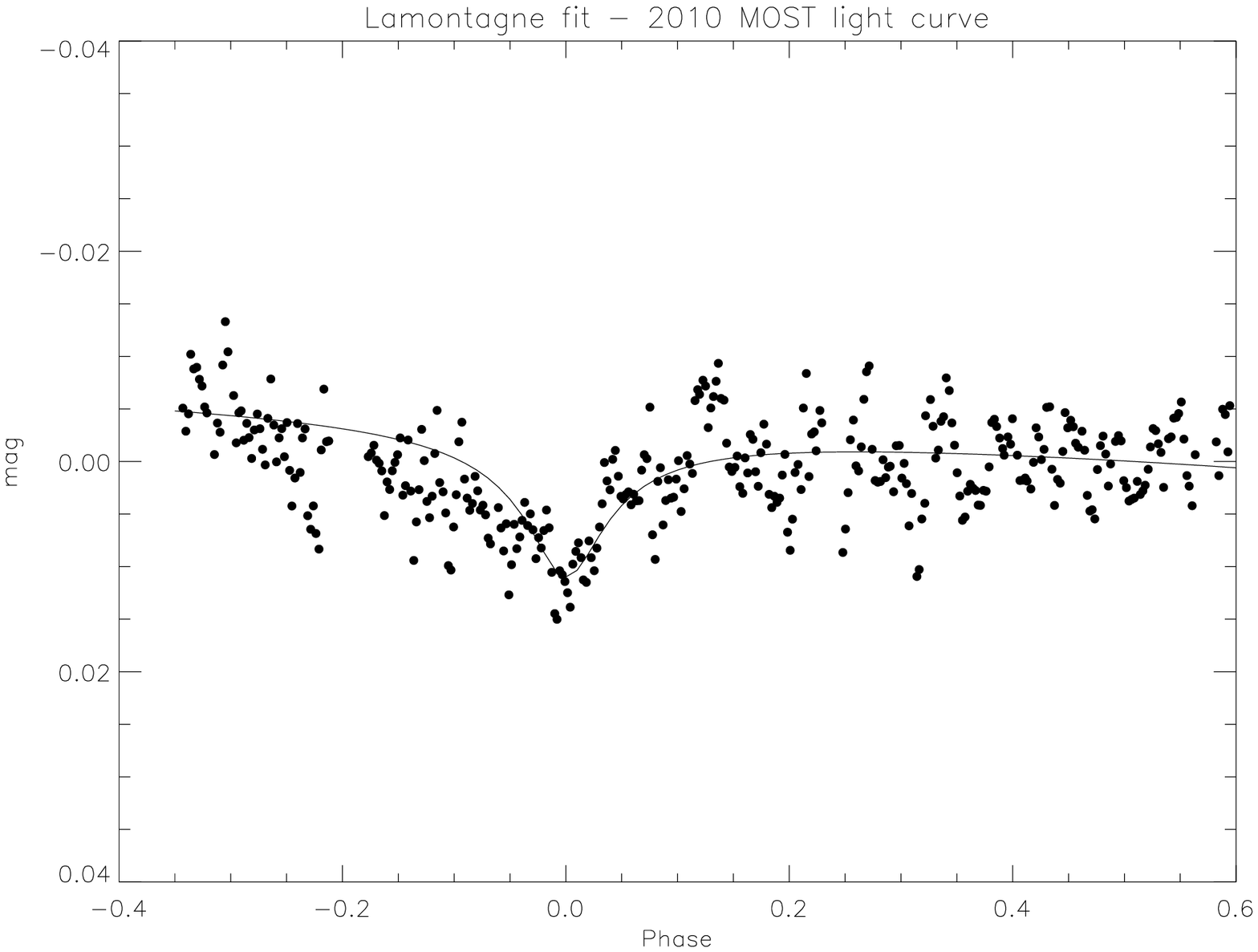}}
\caption{MOST light curve of CV Ser (a) 2009: the most striking feature is the difference between the depth of each eclipse.  Overplotted is the fit
to the data, using the modified \citet{1996AJ....112.2227L} model, as explained in the text.  A typical single-point $2 \sigma$ errorbar (also valid for 2010)
is shown.
(b) 2010: the eclipse seems to be asymmetrical.  
Overplotted is the fit
to the data (as described in the text), using the original Lamontagne model with a linear drift term. The phases were computed using a 29.700d period.}
\end{center}
\label{fig:3}
\end{figure*}

However, there was one small problem trying to apply this model to our data, since it only accounts for constant winds, whereas the changing depth of the eclipses
in CV Ser's light curve clearly indicates that somehow, over one orbital cycle, the parameters of the WR wind have changed.  In order to take this effect
into account, it was necessary to slightly modify the model to include possible variations to the $A$ parameter, which includes most of the important parameters
of the WR wind.  Since little is known about the processes involved in such a change in the wind's structure, there was no reason to assume a particular shape for
the variation of A, so we only included a first-order, time-dependent term (which also seems justified by the relatively short period of time covered by the data).

Because of the instrumental drift, we also allowed $\Delta m_{0}$ (replaced with $B_{0}$ in the formula) to vary linearly with time (with slope $B_{1}$), 
giving the following equation for the modified model:

\begin{equation}
\Delta m = B_{0} + B_{1} t +(A_{0} + A_{1} t) \left( \frac{\pi/2 + \arcsin{\epsilon}}{\sqrt{1-\epsilon^2}} \right),
\end{equation}

\noindent where $t$ is given in units of phase ($[1/P]$).

The corresponding fit is plotted over the light curve in Fig.~3a.  The values of the different parameters are given in Table~\ref{tab:param}.
A Monte-Carlo simulation was used, randomly distributing the errors on the data, in order to evaluate the uncertainties on the fit parameters.

\begin{table}
\caption{Best fit values for the modified Lamontagne model based on the 2009 CV Ser light curve.}
\label{tab:param}
\begin{tabular}{|l|c|}
\hline
Model parameter & Best fit value \\
\hline
$B_{0}$ (mag)   & $-0.0101 \pm 0.0004$ \\
$B_{1}$ (mag/P) & $0.0040 \pm 0.0005$ \\
$A_{0}$ (mag)   & $0.0013 \pm 0.0001$ \\
$A_{1}$ (mag/P) & $0.0008 \pm 0.0001$ \\
$\sin i$        & $0.979 \pm 0.003$ \\
\hline
\end{tabular}
\end{table}

One thing particularly stands out when looking at these values: the ratio $A_{1}P/A_{0} \approx 0.62 \pm 0.12$, which means that the $A$ parameter, 
which was considered
to be constant in the initial model, increases by about 62\% over one complete orbital cycle (29.704d, \citealt{1996RMxAC...5..100N}).  
But how does this affect the values of the physical parameters
of the WR wind?  We must first look at each parameter contained in $A$ individually: the ionization level ($\alpha$), the terminal velocity ($v_{\infty}$),
the intensity ratio ($I_{WR}/I_{O}$) and the mass-loss rate ($\dot{M}$).

The first three parameters all should affect the spectra of the system.  However, as it will be discussed in subsection~\ref{ss:spec}, no such changes were detected.

Therefore, we conclude that only the mass-loss rate varies significantly and its variation is of the order of 62\% over about 30 days, which is unprecedented in any WR star.  
Using the same (constant)
values of $\alpha = 0.5$, $v_{\infty} = 1890 \textrm{km/s}$ and $I_{WR}/I_{O} = 0.69$ as in \citet{1996AJ....112.2227L}, we obtain a value of 
$\dot{M} = 3.5 \times 10^{-6} M_{\odot}y^{-1}$ 
for the first eclipse, and
$\dot{M} = 5.7 \times 10^{-6} M_{\odot}y^{-1}$ for the second eclipse.

Next, the fit was subtracted from the light curve and the residuals were analysed to see if there were any signs of random variations due to clumping.
No significant phase-dependent change in the level of random variability was detected, making a link to clumping unlikely.

The Fourier analysis of the residuals (using period04, \citealt{2005CoAst.146...53L}) yielded a significant peak with a $1.18 \pm 0.03$ c/d frequency 
(see Fig.~4a).  This frequency
was ruled out as an instrumental effect and therefore seemed to be intrinsic to the system.  Time-frequency analysis (Fig.~5a) provided a
clearer picture: both a $\sim$ 1.18 c/d frequency and a broad $\sim$ 0.6 c/d frequency
appeared intermittently (and alternately).  This signature was very similar to that detected in the time-frequency plot of WR110 \citep{2011ApJ...735...34C}, suggesting the presence of corotating interaction regions (CIRs) in WR113's wind.

\begin{figure*}
\begin{center}
\subfigure[2009 periodogram.]{\includegraphics[width=3.4in]{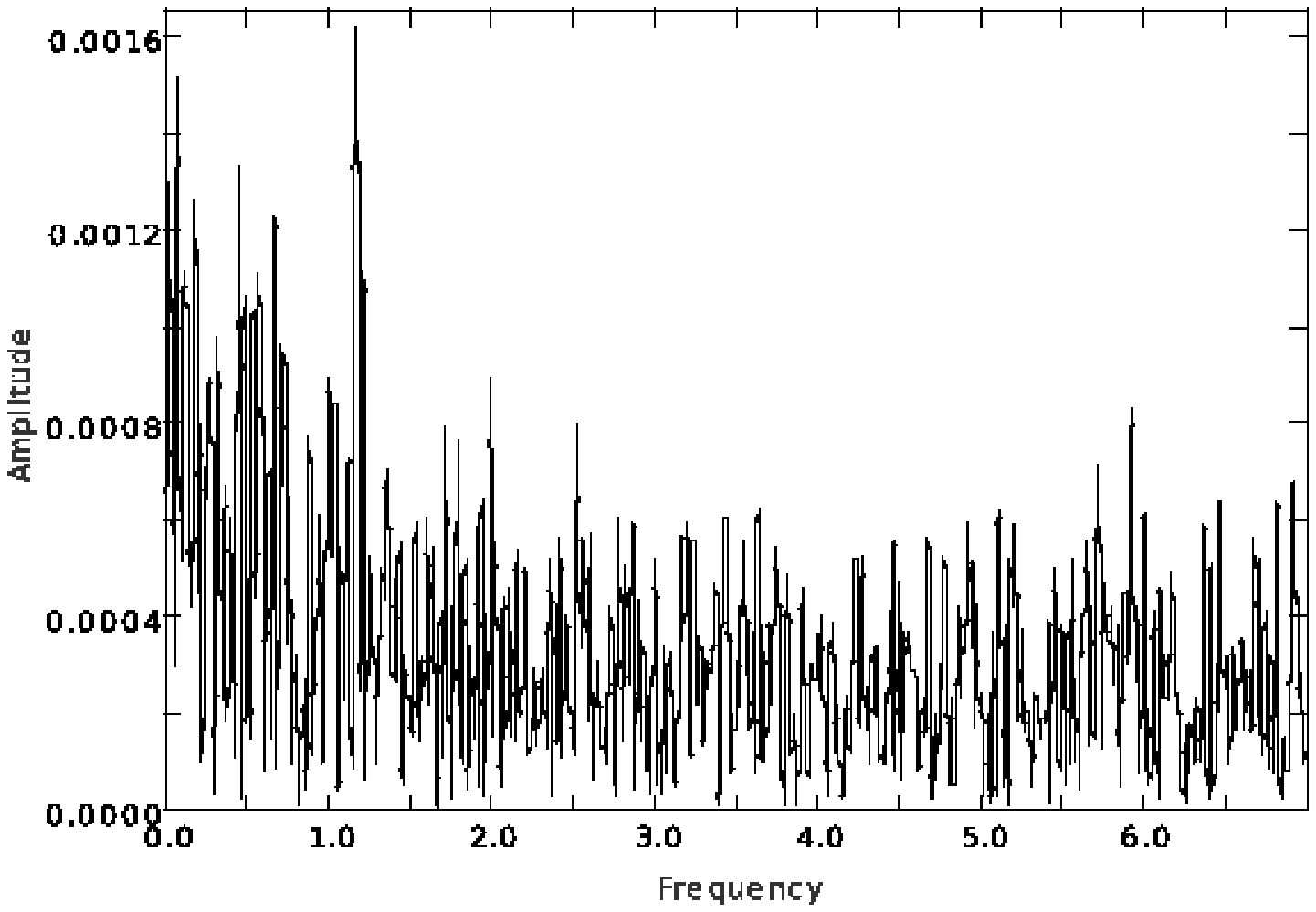}}
\subfigure[2010 periodogram.]{\includegraphics[width=3.4in]{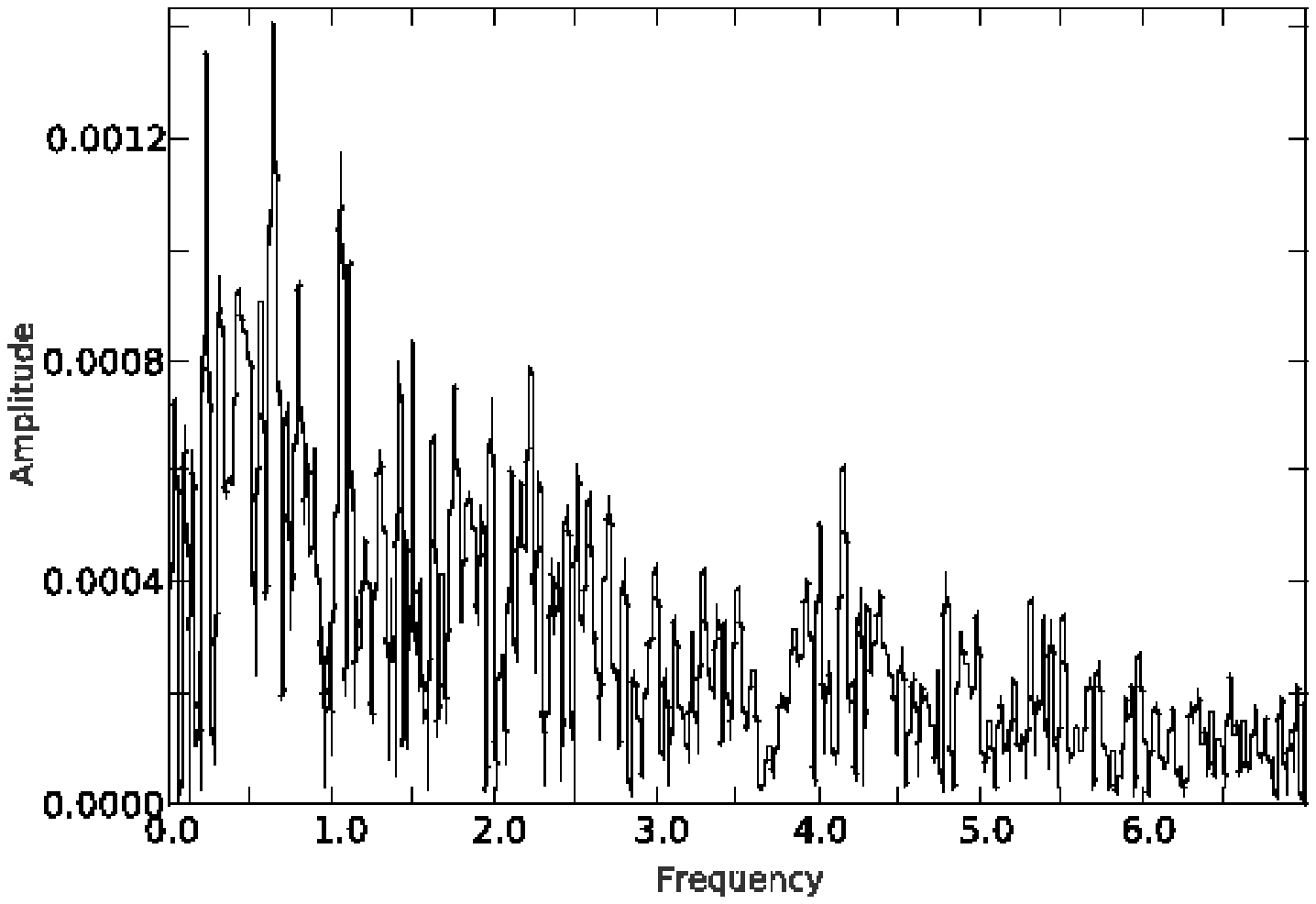}}
\caption{Periodogram of the 2009 MOST light curve residuals (a): a peak can be found at a frequency of 1.18 c/d.
Periodogram of the 2010 MOST light curve residuals (b): the standout peak is at a frequency of 0.64 c/d.
(The amplitude is in mag and the frequency in c/d for both plots.)}
\end{center}
\label{fig:4}
\end{figure*}

\begin{figure*}
\begin{center}
\subfigure[2009 time-frequency plot.]{\includegraphics[width=3.4in]{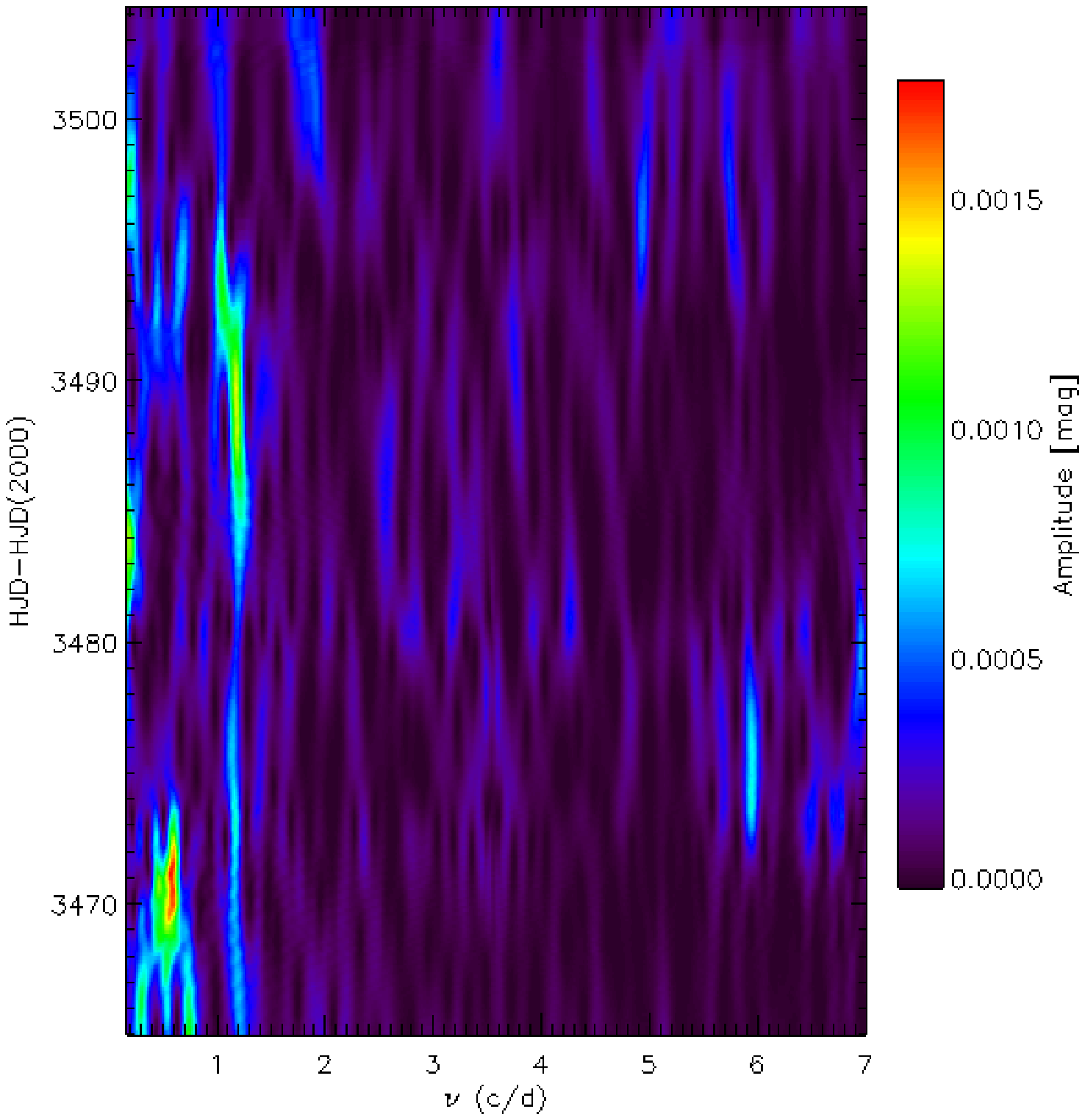}}
\subfigure[2010 time-frequency plot.]{\includegraphics[width=3.4in]{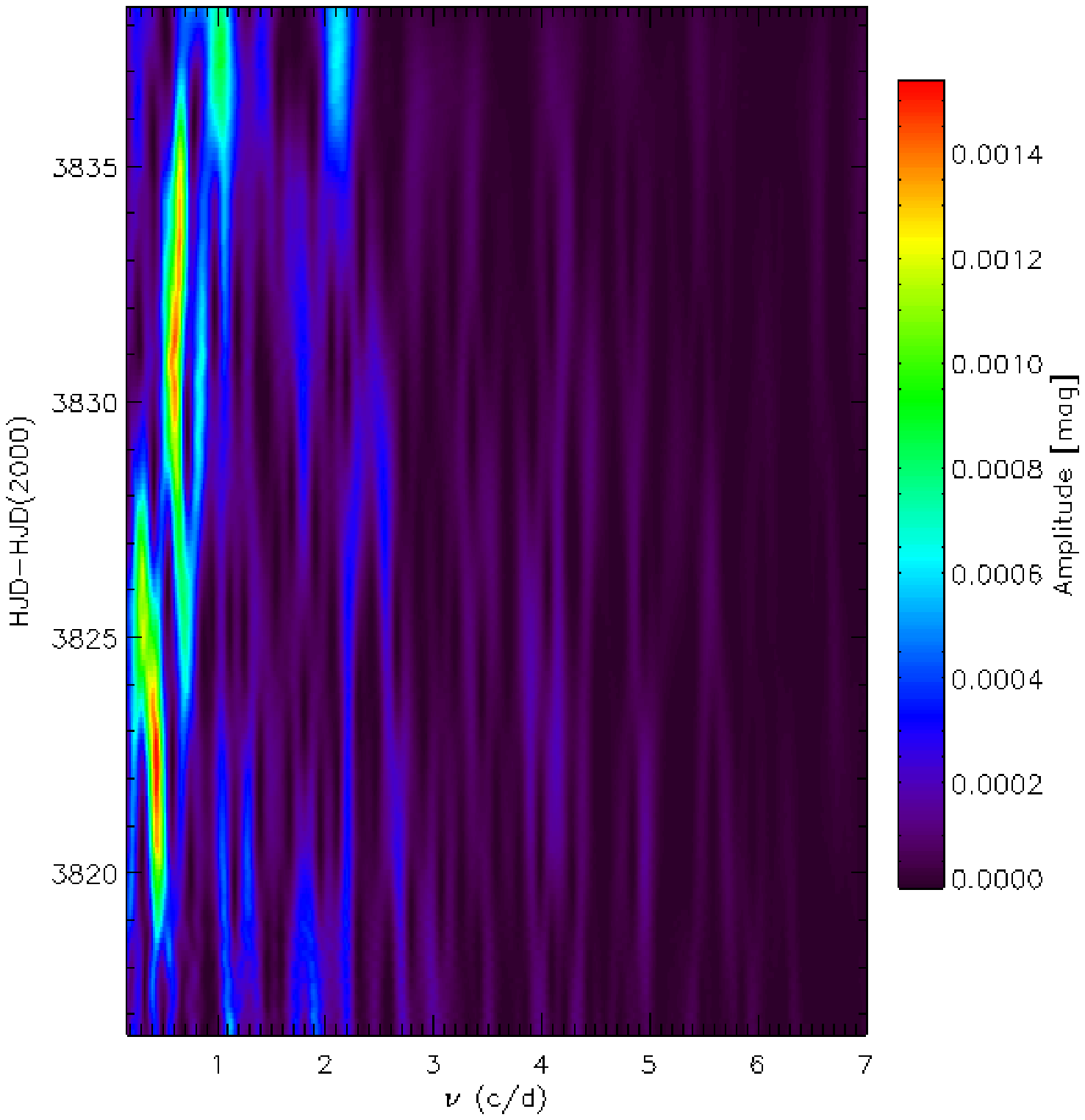}}
\caption{Time-frequency plot of the MOST light curve residuals (a)2009: a CIR-like signature is detected.
(b)2010: a similar signature to that in 2009 is found, strengthening the CIR hypothesis.}
\end{center}
\label{fig:5}
\end{figure*}

The obtained frequencies seem to be coherent with the hypothesis that CIRs may come in pairs (or even higher multiples, e.g. \citealt{1997A&A...327..281K})
per rotation cycle.
Indeed, since one of the frequencies is about twice the other, as is the case for WR110, it seems plausible that the lower frequency is associated with the star's
rotation period, whereas the higher one is due to the presence of two CIR arms.  However, in order to be able to conclude that the star rotates at a frequency of
$\sim$ 0.6 c/d, it would be preferable to see if this frequency is steady over longer periods of time.

\subsection{MOST 2010 Dataset}

The 2010 MOST light curve was binned exactly like that from 2009, in order to be able to make an easier comparison between both light curves.  It consists of 27
days of continuous coverage, spanning across one eclipse (Fig.~3b).

The eclipse shown in this light curve seems to be asymmetrical, a shape somewhat reminiscent of the light curve published by \citet{1963ApJ...137.1080H}, although
it is much shallower.  The same extra absorption during ingress is present.  One hypothesis could be that this additional absorption might be due to a dust event.
However, it isn't clear whether this depression is actually statistically significant or not, since it seems to be quite weak.  In order to test its significance,
the standard Lamontagne et al. model was used to fit the eclipse, once using the entire light curve, and once using only the second half of the eclipse to avoid
including this depression.  Both fits were done using MPFIT (a robust non-linear least squares curve fitting routine based on the Levenberg-Marquardt algorithm, 
\citealt{2009ASPC..411..251M}) and in both cases, 
the value of $\sin\ i$ was fixed to that found in 2009.  

The original model (i.e. with constant $A$ parameter and linear instrumental drift) was used instead of the modified one 
presented earlier in this article since there was only one eclipse in the 2010 light curve.  Indeed, outside of the eclipses, a linear variation of the $A$ 
parameter induced an effect similar to a simple linear drift, since both parameters are correlated (this was made obvious by the Monte-Carlo simulation
used to determine the errors on the fit parameters for the previous dataset), 
so in the case of the 2009 light curve, the only thing making it possible to constrain the value
of that variation was the presence of 2 eclipses.  Therefore, this impossibility of disentangling the linear drift and the variation of $A$ renders any attempt of
evaluating $A_{1}$ futile.  $A$ can only be found accurately during the eclipse, and the value of the instrumental linear drift is not relevant 
(since it cannot be realistically constrained).  

Within the errors, both fits yielded the same values for the model parameters, thus suggesting that the asymmetry is dominated by a random
artefact rather than a truly significant feature of the light curve.  The best fit values of the standard Lamontagne model (including
a parameter to describe once again the instrumental linear drift) are shown in Table~\ref{tab:param2}.   

Deriving the mass-loss rate during the eclipse 
from the fitted value of the $A$ parameter, as for the 2009 data, one gets a value of $\dot{M} = 2.8 \times 10^{-6} M_{\odot}y^{-1}$,
which is the value of the mass-loss rate during the eclipse.  This value is slightly inferior to the mass-loss rate observed during the first eclipse of 2009
(and quite inferior to the rate observed during the second eclipse),
suggesting that this quantity might vary considerably with time in the case of CV Ser (as seen in \citealt{1963ApJ...137.1080H, 1970ApJ...160L.185K}, etc.).


\begin{table}
\caption{Best fit values for the Lamontagne model based on the 2010 CV Ser light curve.}
\label{tab:param2}
\begin{tabular}{|l|c|}
\hline
Model parameter & Best fit value \\
\hline
$B_{0}$ (mag)       & $-0.0040 \pm 0.0003$ \\
$B_{1}$ (mag/P)     & $0.0058 \pm 0.0006$ \\
$A$ (constant, mag) & $0.00104 \pm 0.00005$ \\
$\sin i$ (fixed)    & $0.979$ \\
\hline
\end{tabular}
\end{table}


The fit was then subtracted from the light curve in an effort to try to isolate the cause of the asymmetry (possibly a dust event), but it is not clear from 
the residuals plot
(Fig.~6) that the depression is actually anything but noise (because it is much shallower than the surrounding variability).

\begin{figure}
\includegraphics[width=3.4in]{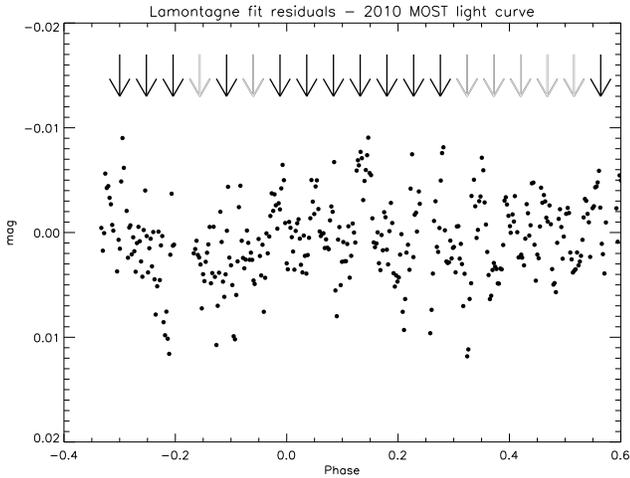}
\caption{2010 MOST light curve residuals. The arrows correspond approximately to the inferred rotation period. Black arrows show peaks in the residuals,
possibly linked to CIRs, whereas no such peaks are found where there are grey arrows (it should be recalled that the CIR phenomenon is expected to be 
cyclical and not strictly periodic). Similar peaks are found in the 2009 residuals.}
\label{fig:6}
\end{figure}

As in the case of the 2009 light curve, no significant signs of random phase-dependent variations were found.  The Fourier analysis of the residuals 
(Fig.~4b) yielded a 
$0.64 \pm 0.03$ c/d peak, coherent with one of the frequencies found in 2009.  
The time-frequency plot (Fig.~5b) shows the same type of signature as the 2009
light curve: two main frequencies (around 0.64 c/d and 1.1 c/d) come and go during the 27-day time series.  The same frequencies were found the previous year,
which is coherent with the CIR scenario, since the apparent frequencies should be modulated by the rotation rate.  Therefore, this suggests a 
rotation period of 1.6d for the WR star according to the argument previously stated,
thus giving a ratio of $v_{rot}/v_{crit} \sim 22\%$, which is relatively high for a WR, \citealt{1998MNRAS.296.1072H}
($v_{crit}$ is calculated using $R_{*} = 5 R_{\odot}$, as in \citealt{1996AJ....112.2227L}).  
It should also be mentioned that a lower frequency is found approximately around
the time of the apparent depression in the light curve (lower left of Fig.~5b).  
It lasts for about 6 days and corresponds to a period of about 3-4 days.  However, it is hard to determine
whether it is linked to the depression or not.



\subsection{Spectroscopy}\label{ss:spec}

One of the goals of this spectroscopic campaign was to solve
the orbit and determine whether it is circular (e.g. \citealt{1981ApJ...245..195M}) or eccentric \citep{1996RMxAC...5..100N}.  
The radial velocities were obtained using the FXCOR routine
in IRAF over the entire range of the DAO spectra, since tests using different ranges show that the excess emission in C{\sc iii} $\lambda5696$ 
and the O star's absorption lines do not perturb significantly the obtained WR RVs. The same range was used with the OMM spectra for consistency.
Fig.~7 shows the relative RV plot, as well as the best circular fit.  Unfortunately, the O star's absorption features being negligible and very weak 
in comparison to the strong, broad emission lines of the WR star, it was not possible to establish that star's
orbit and Fig.~7 only shows the WR orbit.

\begin{figure}
\includegraphics[width=3.4in]{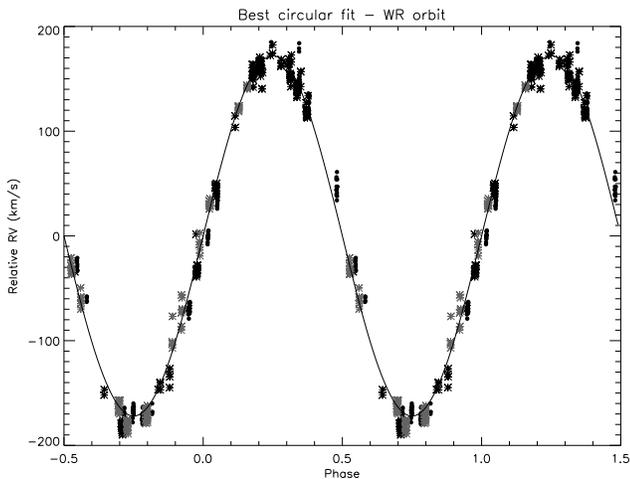}
\caption{Relative radial velocities of the WR component according to phase. The fitted circular orbit is 
overplotted. The dots represent the DAO spectra while the asterisks represent the OMM data (black for 2009, grey for 2010).}
\label{fig:7}
\end{figure}

A first elliptical fit was done using a 29.704d period as in \citet{1996RMxAC...5..100N}.  However, unlike in that article, 
we obtain an orbit which is indistinguishable from circular.  Hence, a circular fit was then applied, and in conjunction with data from \citet{1981ApJ...245..195M},
we obtain a revised period of 29.700d. The obtained value of $K$ is also somewhat larger than previously reported values 
(e.g. \citealt{1981ApJ...245..195M, 1996RMxAC...5..100N}).
The parameters of both fits are shown in Table~\ref{tab:orbit}.

\begin{table}
\caption{Best fit values for the WR orbit based on the current data only.}
\label{tab:orbit}
\begin{tabular}{|l|c|c|}
\hline
Orbital parameter & Elliptical fit & Circular fit \\
\hline
$T_{0}$ (HJD)       & $2455003 \pm 1$ & -                     \\
$E$ (HJD)           & -               & $2455012.6 \pm 0.1$   \\
$K$ (km/s)          & $173 \pm 1$     & $172 \pm 1$           \\
$e$                 & $0.02 \pm 0.02$ & 0 (fixed)             \\
$\omega (^{\circ})$ & $330 \pm 10$    & -                     \\
$P$ (d)             & 29.704 (fixed)  & $29.700 \pm 0.001$    \\
\hline
\end{tabular}
\end{table}

A closer look at the individual spectra reveals considerable spectral variability.  There are different sources of variability which can be distinguished from
one another.  First, there is short-term variability due to the clumped material in the WR wind (including the wind collision zone).  
This is seen as small, variable subpeaks atop some emission lines. Then, some emission lines (C{\sc iii} $\lambda5696$ 
in particular) show excess emission coming from the wind collision zone between the WR and O components.  As the shock cone
orbits with the stars, the excess shifts from one side to the other of the emission line, with a quarter-phase delay.  This effect will be studied further
below.

Finally, it is well known that some emission lines eclipse while others do not (e.g. \citealt{1972SvA....15..955C, 1972PASP...84..772C}).
This behavior is exhibited in the spectra obtained for this study, for instance in the case of C{\sc iii} $\lambda4650$ (as illustrated in 
Fig.~8, this line eclipses while another line, He{\sc ii} $\lambda4686$, does not).  
Even though some lines eclipse, it appears obvious that there is no
dramatic increase or decrease in their EW outside of these eclipses (as shown in Fig.~8 for C{\sc iii} $\lambda4650$).
Therefore, the ratio of the intensities of the O star and the WR star must be quite constant, since e.g. if $I_{WR}/I_{O}$ were
to increase, the emission lines of the WR star would be diluted by the O star's continuum and would have lower equivalent widths.  Thus, we conclude that
$I_{WR}/I_{O}$ does not contribute significantly to the variation of the $A$ parameter, as discussed in section~\ref{ss:2009phot}.


\begin{figure}
\includegraphics[width=3.4in]{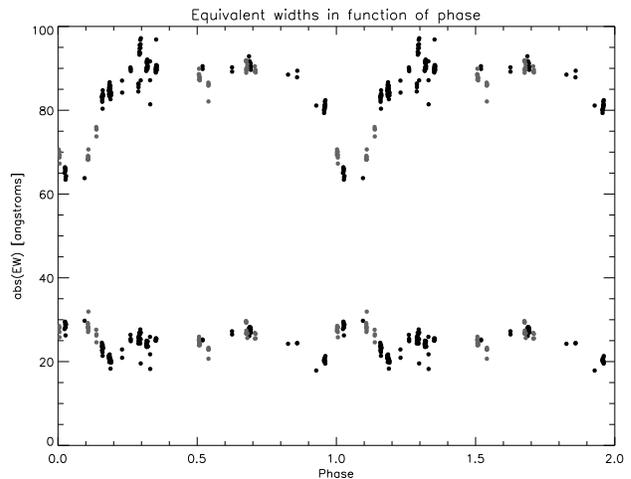}
\caption{Equivalent width of the C{\sc iii} $\lambda4650$ (top) and He{\sc ii} $\lambda4686$ (bottom) lines as a function of phase, 
based on OMM (2009 = black, 2010 = grey) spectra. A double gaussian fit was applied in order to deblend both lines, resulting in some
minor artefacts.}
\label{fig:8}
\end{figure}

Another wind parameter that was monitored is the terminal velocity.  In order to trace any possible variation of $v_{\infty}$, it is best to use a low-ionization
line formed far out in the wind.  In this case, He{\sc i} $\lambda5876$ was used, but no significant variation was found.  Thus, $v_{\infty}$ cannot
account for the variation of the $A$ parameter either.

Finally, following the study of the equivalent widths of individual lines, one can check the ratio between lines of different ionization
levels, to see if the spectral type changes with time.  Two important lines in the determination of WC subtypes are C{\sc iii} $\lambda5696$ 
and C{\sc iv} $\lambda5808$.  The ratio of their equivalent widths (including the excess emission in C{\sc iii} $\lambda5696$ coming from
the wind collision zone, which has constant flux) is fairly constant, 
implying a constant spectral type and, therefore, ionization level.  Consequently, as discussed previously, we conclude that
the only wind parameter that plays a role in the variation of the depth of the eclipses in 2009 is the mass-loss rate.

As mentioned previously, excess emission from the wind-collision zone may offer a lot of information.  In this case, we used the 
model of \citet{1997PASP..109..504L} to study the excess emission in the C{\sc iii} 5696 line. 

The L\"{u}hrs model considers the wind collision zone, which takes the form of a shock cone (with a rounded head).  
This corresponds to the surface where the winds of both stars collide
with equal and opposite momentum components (although more emission may come from the WR side of the shock). 
According to the behavior of the excess emission, it is possible to
determine $\theta$, the half-opening angle of the cone, $\delta\theta$, the thickness of the cone, $\delta\phi$, the phase delay induced by the Coriolis force
acting upon the shock cone as both stars orbit, $v_{strm}$, the streaming velocity of the material as it flows along the cone and $i$, the orbital inclination.

While the L\"{u}hrs model requires an intricate fit of each individual spectrum, there also exists an integral form, as found in \citet{2000MNRAS.318..402H}.
This form only requires the RV and FWHM (full width at half maximum) of the excess emission as a function of phase.  The two main equations are the following :

\begin{equation}\label{eq:fw}
   FWHM_{ex} = C_{1} + 2~v_{strm}~\sin\theta \sqrt{1 - \sin^{2}i\ \cos^{2}(\phi - \delta\phi)}
\end{equation}

\noindent and

\begin{equation}\label{eq:rv}
   RV_{ex} = C_{2} + v_{strm}~\cos\theta\ \sin i\ \cos(\phi - \delta\phi) .
\end{equation}

This form of the model does not yield the thickness of the cone ($\delta\theta$).

In addition to studying the wind collision zone, which in itself is an interesting phenomenon, one might be tempted to try to find proof for the variation of
${\dot{M}}_{WR}$ by applying this model, since there exists a relationship between the half-opening angle of the shock cone and the ratio of the wind momenta.
\citet{1995IAUS..163..495U} finds the following formula :

\begin{equation}
   \theta (rad) \simeq 2.1\ \left(1-\frac{\eta^{2/5}}{4}\right) \eta^{1/3},
\end{equation}

\noindent where $\eta = \frac{{\dot{M}}_{O} v_{\infty,O}}{{\dot{M}}_{WR} v_{\infty,WR}}$ and is comprised between $10^{-4}$ and 1.  If the half-opening
angle of the shock cone can accurately map out the variations of the WR's mass-loss rate, such a spectral analysis could
prove useful in order to monitor a system like CV Ser, where $\dot{M}$ seems to vary quite a bit.

Fig.~9 shows the C{\sc iii} $\lambda5696$ 
line in the WR reference frame.  It is quite obvious here that the overlying excess varies, but in order
to be able to isolate it, a minimum profile must be subtracted, which should correspond to the unperturbed emission line.  Here, the line seems a little slanted,
so the base profile (shown in Fig.~9) was chosen to be a tilted flat top 
(in order to fit both the wings and the unperturbed edges of the top of the line).
\footnote{Although the choice of such a profile may seem a bit \textit{ad hoc}, similar profiles have been found in real
spectra (e.g. in the single WC8 star WR135, \citealt{1999ApJ...514..909L}).}  
Once subtracted, it is easier to see how the excess varies with phase
(Fig.~10).

\begin{figure*}
\includegraphics[width=5.0in]{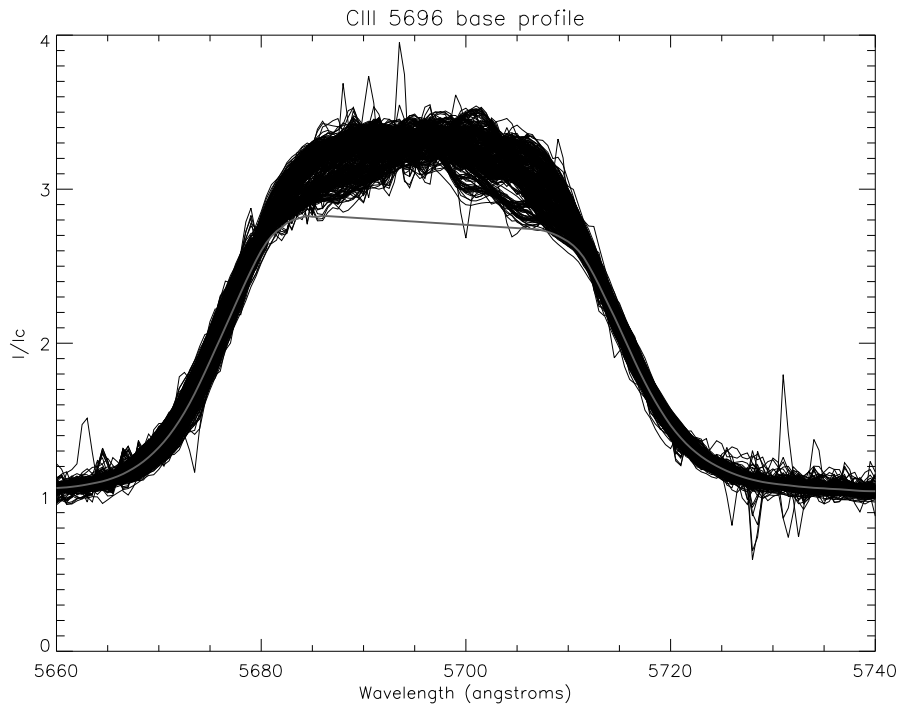}
\caption{Base profile (grey line), which is to be subtracted from the C{\sc iii} $\lambda5696$ line to isolate the excess emission.  
The C{\sc iii}
$\lambda5696$ profiles from both the DAO (2009) and OMM (2009, 2010) data are shown, shifted in the WR reference frame, with their base profile.}
\label{fig:9}
\end{figure*}

\begin{figure}
\includegraphics[width=2.4in]{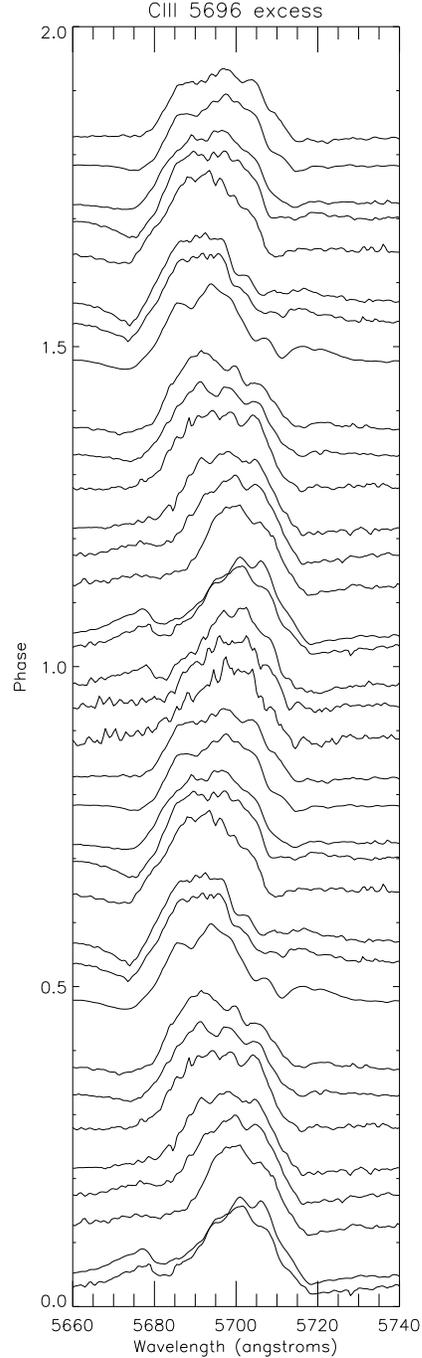}
\caption{Excess emission in C{\sc iii} $\lambda5696$ varying with phase along the vertical axis. Combined data from OMM (2009, 2010) and DAO (2009), binned by
phase (bins of 0.05).}
\label{fig:10}
\end{figure}

The L\"{u}hrs analysis was first performed by calculating the first moment (mean) of the excess for the RV and roughly 
twice the square root of the second centered moment (standard deviation)
for the FWHM.  One difficulty encountered using this technique is that the result varied greatly depending upon what integration bounds were used.  Consequently,
this technique was not reliable enough, using the available data, and had to be abandoned and replaced by a simpler one - actually measuring the 
FWHM and using the midpoint at half maximum to compute the radial velocities.  

To do so, the OMM and DAO spectra were combined into a single dataset, 
which was then binned to get rid of short-term stochastic effects as well as diminish the noise.
Each individual bin covers 1/20 of the orbital period.
The inclination angle, $i$, was fixed to the value found using the photometry, since it is not well constrained by 
Eqs.~\ref{eq:fw} and~\ref{eq:rv} in the presence of instrumental noise.  
The results are shown in Fig.~11.

\begin{figure}
\includegraphics[width=3.4in]{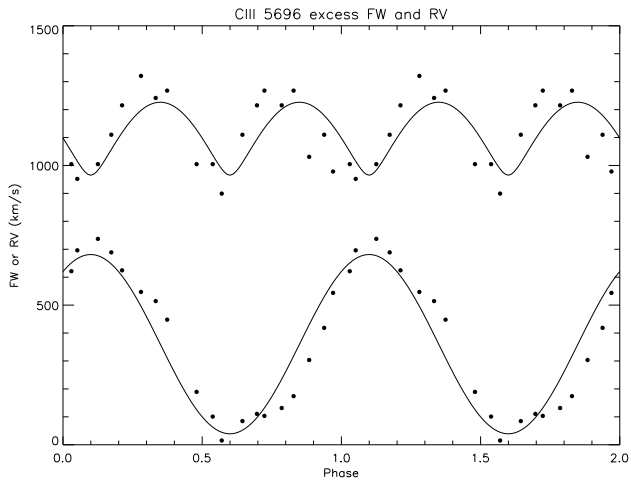}
\caption{FW (top) and RV (bottom) of the excess emission as a function of phase. Each dot correspond to a particular phase bin (the OMM and DAO data were combined
together for this analysis).}
\label{fig:11}
\end{figure}



The fit was first performed without letting $\theta$ vary with time.  The results are presented in Table~\ref{tab:luhrs}. Unfortunately, since the dependence of
$\theta$ on $\eta$ is rather weak, the slight effect it would have on the data is too subtle to be well constrained; therefore it was not possible to highlight
the effects of the varying mass-loss rate on the shock cone.  Also, the obtained values are somewhat surprisingly low and far from those found in the previous study
of \citet{2000ASPC..204..295A}.  Using the \citet{1995IAUS..163..495U} formula, an $\eta$ parameter of about $0.012$ is found, whereas using 
\citet{2005A&A...436.1049M} (to obtain reasonable stellar parameters according to the O star's spectral type) and \cite{2001A&A...369..574V} (to obtain 
$\dot{M}_{O}$ and $v_{\infty, O}$ in function of those stellar parameters), 
we expect to find a value of about $0.08$. The two obtained values are a little less than an order of magnitude apart, which seems fairly reasonable given
the many possible sources of error for such a calculation.

\begin{table}
\caption{Best fit values for the L\"{u}hrs analysis (based on the DAO spectra).}
\label{tab:luhrs}
\begin{tabular}{|l|c|}
\hline
Model parameter & Best fit value \\
\hline
$v_{strm}$ (km/s)  & $366 \pm 18$ \\
$\theta$ (deg)     & $27 \pm 3$ \\
$\delta\phi$ (deg) & $36 \pm 2$ \\
\hline
\end{tabular}
\end{table}

Finally, one of the secondary aims of the spectroscopic part of this study was to confirm the spectral types of both orbital companions.  In order to disentangle
the spectra of both stars, the "shift and add" method described in \citet{2002ApJ...577..409D} was to be used.  This method is iterative and uses the orbits to
"shift" the spectra to the reference frame of the WR star.  Then, the spectra are "co-added" and the resulting compound spectrum is then shifted back to our reference 
frame to be subtracted from the initial spectra, revealing the spectra of the O star.  After a few iterations, it is possible to get the spectrum of each star.
Since the orbit of the O star was not obtained in this study, we used the orbit found in a previous study with similar ephemeris \citep{1981ApJ...245..195M}.  
The method was applied over 10 iterations.  However, the obtained O spectrum is heavily contaminated by
artefacts due to the strong WR emission lines.  This may be partly due to a slightly wrong orbital solution.  In any case, it is impossible to confirm the spectral
classification of the O companion based on these results.


However, the WR spectrum does confirm the previous WC8d classification (the `d' denotes the detection of dust in the IR), 
since it corresponds to the features found in \citet{1990ApJ...358..229S}.  In particular, the 
C{\sc iv} $\lambda$5808/C{\sc iii} $\lambda$5696 peak flux ratio is about 0.7, which lies well within the 0.5-2.0 range associated with the WC8 spectral type
(this ratio goes up to $\sim 0.8$ if the excess emission produced in the wind collision zone is removed).

It should also be mentioned that no attempt was made to detect the hypothetical CIRs in the spectroscopic data, as the signal is complicated by the presence of
excess emission from the wind collision zone.

\section{Discussion and Conclusions}\label{sec:disc}

After having eliminated the other possible causes for the change in eclipse depth observed in the 2009 MOST photometry, the deduced 62\% increase in $\dot{M}$
over one 29.704d orbital period is a truly remarkable finding, unprecedented for WR stars.  Even more intriguing is the fact that the following year, the mass-loss
rate had gone back to a lower value, suggesting that it may vary considerably over short to long timescales.  The derived values of $\dot{M}$ are also fairly
low for a WR star.  According to \citet{2002A&A...392..653C}, variations in wind density should cause the spectral type to change, but this effect is not observed
here, adding to the mystery, although these results are not necessarily incompatible since it is noted that the dependence on the wind density is weaker
for late-type WC stars (WC8 and WC9).  
In any case, this definitely constitutes a challenge to both theorists and observers, since there is a dire need for a theoretical explanation
as to what could drive such an impressive variation of the mass-loss rate, while constant monitoring would be required in order to get a better idea of the long-term
behaviour of this parameter. One alternate explanation could conceivably be 
the presence of thin-shell instabilities in the wind collision zone. However, this will only affect the light curve dips stochastically, not systematically.
In any case, it is not so much the details of the wind collision which determine the eclipse's shape and depth; rather it is the global wind
of the WR component in which the O star orbits.

As for the primary goal of this study, it was not possible to systematically link the stochastic photometric variability to the clumping phenomenon.  There was no
clear phase dependence of the scatter of the light curves.  The variability is then probably due to noise (the combination of the unquantifiable absorption due
to clumping and other stochastic phenomena), but also to CIRs in the WR wind, as the analysis of the
light curve residuals suggests.  The repeated signature (both in 2009 and 2010) offers a rather robust clue towards that hypothesis.  This finding is very 
interesting in the actual context of CIRs in WR stars, since more and more CIR candidates are showing up
(e.g. \citealt{2009ApJ...698.1951S}, \citealt{2011ApJ...735...34C}, \citealt{2011ApJ...736..140C}).  Even though they have mainly been
considered as an exception up until now, CIRs in WR winds might just prove to be the rule after all.  Further studies will be necessary in order to find cyclical 
spectroscopic evidence supporting this possibility.  However, their origin remains mysterious, especially since
no magnetic fields have been detected in the extended atmospheres of WR stars yet.  Nevertheless, this scenario should not be ruled out since at the base of the
winds, where the CIRs originate, the magnetic field might have much higher values than our current detection limits but still is not detected simply because this
region is obscured by the dense wind (or perhaps due to cancellation in small dipole loops as seen on sunspots). Non radial pulsations (NRP) or starspots have also
been suggested as possible causes for this phenomenon \citep{1996ApJ...462..469C}.

Finally, what was initially thought to be a possible dust event during the 2010 observations cannot conclusively be determined as statistically significant.
However, the perceived asymmetry in the light curve is reminiscent of the model presented by \citet{1998A&A...329..199V}.  Although this model was used
to characterize the condensation of dust clouds and its effect on a light curve, and as such is very unlikely linked to these observations (since 
such a dust event should create a much deeper dip in the light curve and the probability of it occurring exactly at the same time as the eclipse is rather low), 
it still gives way to a very important question : what role can dust play in CV Ser's bizarre behaviour? Could a variation of the quantity of dust
persistently produced by the WR star affect the light curve in the same way it has been inferred that a variation of the total mass-loss rate might? 
In order for such an explanation to arise, one must first postulate that the dust distribution in the wind is the same as the free electron and hence
overall density distribution, in order
to preserve the system's geometrical properties and to reproduce the same eclipse profile found in the light curve. As to whether this assumption makes physical
sense, it is hard to determine whether the dust could be produced (or maintained) following such a spatial distribution, since the production of dust in
late WC stars is not yet well understood. However, it is fairly straightforward to calculate what quantity of dust would be needed to produce a similar
effect to that of an increased mass-loss rate on the light curve. Assuming Mie scattering, we get a cross-section, for one grain of average
radius $a$, of $\sigma_{d} = Q \pi a^{2}$, where 
$Q$ is the cross-section efficiency. Choosing $a$ to be $\sim 0.1$ $\mu$m (as in WR140: \citealt{2003ApJ...596.1295M}), we find that
$Q \sim 2$ for wavelengths of the order of 5000 \AA , giving a value of $\sigma_{d} \sim 6.3 \times 10^{-10}$ cm$^{2}$. However, using a density inside the dust
grains of $\rho_{g} = 2$ g/cm$^{3}$, we find that each grain is composed of roughly $4 \times 10^{8}$ carbon atoms. Since the most common carbon ion in the WR
wind is C{\sc iii}, for each atom of carbon we should expect to find 2 free electrons. Then, the Thomson cross-section of the number of free electrons needed to 
combine with $4 \times 10^{8}$ carbon atoms to produce one grain
is $\sigma_{e, tot} = 8 \times 10^{8} \times \sigma_{e} = 5.3 \times 10^{-16}$ cm$^{2}$. Therefore, scattering by the equivalent quantity of recombined
carbon in the form of a dust grain is about $10^{6}$ times more efficient, therefore an appreciable increase in the depth of the eclipse could be caused
by recombination of a negligibly small quantity of carbon and its condensation in the form of dust. Possible support for this is seen in the UBV light curve
of CV Ser obtained from 1984 to 1994 by Dzhapiashvili (Anthokhin, priv. com.) in which the eclipse depth varies strongly with wavelength.
Therefore, if dust plays any role in these varying eclipse depths, the 62\% value for the increase of $\dot{M}$ 
over an orbital period in the 2009 data obtained in this study should probably be considered as an upper limit.
Unfortunately, not much more can be deduced about the production of dust in CV Ser, except that it remains a very interesting phenomenon and should be 
further studied.  Indeed, with its troubled history and intriguing behaviour, CV Ser might prove to be a key system for understanding the production of dust 
in WC+O binaries. It does not appear clear whether this process is necessarily due to the wind collision zone or if it originates in wind shocks and is then 
intrinsic to the WR component. 

In conclusion, this work possibly raises more questions than it answers, but we conclude without a doubt that CV Ser is a very important system that might
hold the answer to old problems.  Hopefully, our findings will motivate the community to take a deeper look into this remarkable object in the years to come.

\section*{Acknowledgments}

A.D.-U. would like to acknowledge the help and support of his Universit\'{e} de Montr\'{e}al colleagues, in particular N. St-Louis, R. Fahed, 
A. de la Chevroti\`{e}re and S. Desforges, as well as D. Souti\`{e}re and S. P\'{e}loquin for their help in acquiring part of the data.  
The Natural Sciences and Engineering Research Council of Canada supports the research of D.B.G., J.M.M., A.F.J.M. and S.M.R., while A.F.J.M. is
also supported by Le Fonds qu\'{e}b\'{e}cois de la recherche sur la nature et les technologies.  R.K. and W.W.W. are supported by the Austrian Space
Agency and the Austrian Science Fund. A.-N.C. gratefully acknowledges support from the Chilean Centro de Astrof\'{i}sica FONDAP No. 15010003 and the Chilean
Centro de Excelencia en Astrof\'{i}sica y Tecnolog\'{i}as Afines (CATA) and Comitee Mixto ESO-GOBIERNO DE CHILE.

This research has made use of the SIMBAD database operated at CDS, Strasbourg, France and NASA's Astrophysics Data System (ADS) Bibliographic Services.

Finally, the authors thank the anonymous referee for his insightful comments which have no doubt contributed to making this paper better.

\label{lastpage}


\begin{thebibliography}{99}
\bibitem[\protect\citeauthoryear{Allen et al.}{1972}]{1972A&A....20..333A} Allen D.A., Swings J.P., Harvey P.M.,
1972, A\&A, 20, 333
\bibitem[\protect\citeauthoryear{Antokhin et al.}{2000}]{2000ASPC..204..295A} Antokhin I.I., Hill G.M., Moffat A.F.J., 2000, ASPC, 204, 295
\bibitem[\protect\citeauthoryear{Chen\'{e} et al.}{2011}]{2011ApJ...735...34C} Chen\'{e} A.N., Moffat A.F.J., Cameron C.,
Fahed R., Gamen R.C., Lef\`{e}vre L., Rowe J.F., St-Louis N., Muntean V., De La Chevroti\`{e}re A., Guenther D.B., Kuschnig R., 
Matthews J.M., Rucinski S.M., Sasselov D., Weiss W.W., 2011, ApJ, 735, 34
\bibitem[\protect\citeauthoryear{Chen\'{e} \& St-Louis}{2011}]{2011ApJ...736..140C} Chen\'{e} A.N., St-Louis N., 2011, ApJ, 736, 140
\bibitem[\protect\citeauthoryear{Cherepashchuk}{1972}]{1972SvA....15..955C} Cherepashchuk A.M., 1972, SvA, 15, 955
\bibitem[\protect\citeauthoryear{Cohen et al.}{1975}]{1975A&A....40..291C} Cohen M., Kuhi L.V., Barlow M.J.,
1975, A\&A, 40, 291
\bibitem[\protect\citeauthoryear{Cowley et al.}{1971}]{1971A&A....11..407C} Cowley A.P., Hiltner W.A.,
Berry C., 1971, A\&A, 11, 407
\bibitem[\protect\citeauthoryear{Cowley}{1972}]{1972PASP...84..772C} Cowley A., 1972, PASP, 84, 772
\bibitem[\protect\citeauthoryear{Cranmer \& Owocki}{1996}]{1996ApJ...462..469C} Cranmer S.R., Owocki S.P., 1996, ApJ, 462, 469
\bibitem[\protect\citeauthoryear{Crowther et al.}{2002}]{2002A&A...392..653C} Crowther P.A., Dessart L., Hillier D.J.,
Abbott J.B., Fullerton A.W., 2002, A\&A, 392, 653
\bibitem[\protect\citeauthoryear{Demers et al.}{2002}]{2002ApJ...577..409D} Demers H., Moffat A.F.J.,
Marchenko S.V., Gayley K.G., Morel T., 2002, ApJ, 577, 409
\bibitem[\protect\citeauthoryear{Gaposchkin}{1949}]{1949PZ......7...36G} Gaposchkin S., 1949,
PZ, 7, 36
\bibitem[\protect\citeauthoryear{Harries et al.}{1998}]{1998MNRAS.296.1072H} Harries T.J., Hillier D.J., Howarth I.D., 1998, MNRAS, 296, 1072
\bibitem[\protect\citeauthoryear{Hill et al.}{2000}]{2000MNRAS.318..402H} Hill G.M., Moffat A.F.J., St-Louis N., Bartzakos P.,
2000, MNRAS, 318, 402
\bibitem[\protect\citeauthoryear{Hiltner}{1945}]{1945ApJ...101..356H} Hiltner W.A., 1945, ApJ, 101, 356
\bibitem[\protect\citeauthoryear{Hjellming \& Hiltner}{1963}]{1963ApJ...137.1080H} Hjellming R.M., Hiltner W.A.,
1963, ApJ, 137, 1080
\bibitem[\protect\citeauthoryear{Kaper et al.}{1997}]{1997A&A...327..281K} Kaper L., Heinrichs H.F., Fullerton A.W., 
Ando H., Bjorkman K.S., Gies D.R., Hirata R., Kambe E., McDavid D., Nichols J.S., 1997, A\&A, 327, 281
\bibitem[\protect\citeauthoryear{Kuhi \& Schweizer}{1970}]{1970ApJ...160L.185K} Kuhi L.V., Schweizer F., 1970,
ApJ, 160, 185
\bibitem[\protect\citeauthoryear{Lamontagne et al.}{1996}]{1996AJ....112.2227L} Lamontagne R., Moffat A.F.J., Drissen L.,
Robert C., Matthews J.M., 1996, AJ, 112, 2227
\bibitem[\protect\citeauthoryear{Lenz \& Breger}{2005}]{2005CoAst.146...53L} Lenz P., Breger M., 2005, Communications in Asteroseismology,
146, 53
\bibitem[\protect\citeauthoryear{L\'{e}pine \& Moffat}{1999}]{1999ApJ...514..909L} L\'{e}pine S., Moffat A.F.J, 1999, ApJ, 514, 909
\bibitem[\protect\citeauthoryear{Lipunova}{1985}]{1985Ap&SS.109...57L} Lipunova N.A., 1985, Ap\&SS, 109, 57
\bibitem[\protect\citeauthoryear{L\"{u}hrs}{1997}]{1997PASP..109..504L} L\"{u}hrs S., 1997, PASP, 109, 504
\bibitem[\protect\citeauthoryear{Marchenko et al.}{2003}]{2003ApJ...596.1295M} Marchenko S.V., Moffat A.F.J., Ballereau D., Chauville J.,
Zorec J., Hill G.M., Annuk K., Corral L.J., Demers H., Eenens P.R.J., Panov K.P., Seggewiss W., Thomson J.R., Villar-Sbaffi A., 2003, ApJ, 596, 1295
\bibitem[\protect\citeauthoryear{Markwardt}{2009}]{2009ASPC..411..251M} Markwardt C.B., 2009, ASPC, 411, 251
\bibitem[\protect\citeauthoryear{Martins et al.}{2005}]{2005A&A...436.1049M} Martins F., Schaerer D., Hillier D.J., 2005, A\&A, 436, 1049
\bibitem[\protect\citeauthoryear{Massey \& Niemela}{1981}]{1981ApJ...245..195M} Massey P., Niemela V.S., 1981, ApJ, 245, 195
\bibitem[\protect\citeauthoryear{Morrison \& Wolff}{1972}]{1972PASP...84..635M} Morrison N.D., Wolff S.C., 1972, PASP, 84, 635
\bibitem[\protect\citeauthoryear{Niemela et al.}{1996}]{1996RMxAC...5..100N} Niemela V.S., Morrell N.I., Barba R.H., Bosch G.L., 1996, RMxAC, 5, 100
\bibitem[\protect\citeauthoryear{Pauldrach et al.}{1986}]{1986A&A...164...86P} Pauldrach A., Puls J., Kudritzki R.P., 1986, A\&A, 164, 86
\bibitem[\protect\citeauthoryear{Smith et al.}{1990}]{1990ApJ...358..229S} Smith L.F., Shara M.M., Moffat A.F.J., 1990, ApJ, 358, 229
\bibitem[\protect\citeauthoryear{St-Louis et al.}{2009}]{2009ApJ...698.1951S} St-Louis N., Chen\'{e} A.N., Schnurr O., Nicol M.H., 2009,
ApJ, 698, 1951
\bibitem[\protect\citeauthoryear{Stepie\'{n}}{1970}]{1970AcA....20...13S} Stepie\'{n} K., 1970, AcA, 20, 13
\bibitem[\protect\citeauthoryear{Tucker}{1977}]{1977rpa..book.....T} Tucker W.H., \textit{Radiation processes in astrophysics}, Cambridge MA, MIT Press, 1977.
\bibitem[\protect\citeauthoryear{Usov}{1995}]{1995IAUS..163..495U} Usov V.V., 1995, IAUS, 163, 495
\bibitem[\protect\citeauthoryear{Veen et al.}{1998}]{1998A&A...329..199V} Veen P.M., van Genderen A.M., van der Hucht K.A., Li A.,
Sterken C., Dominik C., 1998, A\&A, 329, 199
\bibitem[\protect\citeauthoryear{Vink et al.}{2001}]{2001A&A...369..574V} Vink J.S., de Koter A., Lamers H.J.G.L.M., 2001, A\&A, 369, 574
\bibitem[\protect\citeauthoryear{Walker et al.}{2003}]{2003PASP..115.1023W} Walker G., Matthews J., Kuschnig R.,
Johnson R., Rucinski S., Pazder J., Burley G., Walker A., Skaret K., Zee R., Grocott S., Carroll K., Sinclair P., Sturgeon D.,
Harron J., 2003, PASP, 115, 1023
\bibitem[\protect\citeauthoryear{Williams et al.}{1977}]{1977Obs...97....76W} Williams P.M., Beattie D.H., Stewart J.M., 1977, The Obs., 97, 76
\bibitem[\protect\citeauthoryear{Williams et al.}{1987}]{1987A&A...182...91W} Williams P.M., van der Hucht K.A., Th\'{e} P.S.,
1987, A\&A, 182, 91
\bibitem[\protect\citeauthoryear{Williams}{1995}]{1995IAUS..163..335W} Williams P.M., 1995, IAUS, 163, 335

\end{thebibliography}
\end{document}